# Federated Learning for Iot/Edge/Fog Computing Systems

**Balqees Talal Hasan[1] and Ali Kadhum Idrees[2] [*]**

[1]*Dept of Computer and Information Engineering, College of Electronics Engineering, Nineveh University, Mosul, Iraq*

[2]*Dept of Information Networks, College of Information Technology, University of Babylon, Babylon, Iraq*

[*]*Corresponding author Emails: **ali.idrees@uobabylon.edu.iq***

**ABSTRACT**

With the help of a new architecture called Edge/Fog (E/F) computing, cloud computing services can now be extended nearer to data generator devices. E/F computing in combination with Deep Learning (DL) is a promisedtechnique that is vastly applied in numerous fields. To train their models, data producers in conventional DL architectures with E/F computing enable them to repeatedly transmit and communicate data with third-party servers, like Edge/Fog or cloud servers. Due to the extensive bandwidth needs, legal issues, and privacy risks, this architecture is frequently impractical. Through a centralized server, the models can be co-trained by FL through distributed clients, including cars, hospitals, and mobile phones, while preserving data localization. As it facilitates group learning and model optimization, FL can therefore be seen as a motivating element in the E/F computing paradigm. Although FL applications in E/F computing environments have been considered in previous studies, FL execution and hurdles in the E/F computing framework have not been thoroughly covered. In order to identify advanced solutions, this chapter will provide a review of the application of FL in E/F computing systems. We think that by doing this chapter, researchers will learn more about how E/F computing and FL enable related concepts and technologies. Some case studies about the implementation of federated learning in E/F computing are being investigated. The open issues and future research directions are introduced.

## 1. INTRODUCTION

The number of Internet of Things (IoT)devices has significantly increased, resulting in a remarkable increase in generated data.In 2025, the number of nodes connected to the Internet is projected to increase to 80 billion and 180 trillion gigabytes (GB) of data will be produced globally(Shaheen et al., 2022).IoT systems typically dependon cloud-based IoT architecture, in which servers on the cloud are used to manage the physical objects.Cloud-based architecture faces several challenges, including security, latency, and bandwidth. In response to these challenges, new technologies likefog and edge computing have emerged, attempting to extend the cloud computing model by pushingstorage, computing, and networking resources significantly nearer to data sources or end-user apps (Buyya & Srirama, 2019).

Smart objects must support some form of artificial intelligence (AI).The success of deep learning (DL) in video and speech applications—two of the most essential IoT services—has made it an extremely important area of research to adapt its models for deployment on edge computing devices with limited resources(Mohammadi et al., 2018).In order to support DL, new hardware has been introduced; typically, an external chip or graphical processing unit (GPU) is introduced for this purpose and is referred to as an AI chip. However, AI chips are costly, making it uneconomical or unaffordable for them to be utilized in a variety of products.Furthermore, as dataset sizes grow, machine learning (ML) models get more complex, with a DNN using millions of parameters (Malekijoo et al., 2021).Besides this to train traditional DL models in edge computing devices, owners of data must regularly shareraw data with external servers such as edge or cloud servers. Because of thelegalization, high bandwidth requirements, and privacy vulnerabilities, this architecture is frequently impractical(Abreha, 2022).As a result, decentralized approaches, like Federated Learning (FL), became an unavoidable substitute for centralized ones(Malekijoo et al., 2021).

In 2016, Google has offered FL as a distributed machine learning (ML)approach to sendprocessing of AI applications into an expanding number of end deviceswhile safeguarding end users' privacy(Ye et al., 2020).Since then, FL has rapidly developed and turned into a popular AI research topic(Yin et al., 2021).Fully distributed training is made possible by FL by splitting the process into two stages: clients-side parallel model updating based on local dataset and server-side global model aggregation(L. Liu et al., 2020).FL is thought to be a successful strategy for utilizing distributed resources to cooperatively train an ML model. FL maintains the decentralized nature of the raw data and keeps it away from a single server or data center while enabling several clients to collaborate to train a model.In FL, only intermediate data may be sent between distributed computing resources; training data transmission is prohibited(J. Liu et al., 2022).FL incorporates techniques from a variety of research areas, including distributed systems, machine learning, and privacy(Li et al., 2021).Applications of FL approach are found in a variety of business sectors, including pharmaceutics, medical AI, telecom, IoT, transportation, and traffic prediction and monitoring(Shaheen et al., 2022).

This chapter is organized as follows. The Federation concept is introduced in Section 2. FL Workflows is presented in section 3. In section 4, Federated Learning vs. Non-Federated Learning is explained. The Scale of Federation is illustrated in section 5. The FL Network Topology is introduced in section 6. The Classification of a Federation are presented and explained in section 7. We look into Federated Learning Aggregation Algorithms in section 8.The fundamentals of cloud, edge, and fog computing are introduced in section 9. Then, enabling Deep Learning at the E/F is presented in section 10. The FL Implementation in E/F Computing is being investigated in section 11. The FL datasets is presented in section 12. Section 13 demonstrates FL applications in E/F computing. Section 14 discusses the open issues and future FL research directions. Finally,Section 15 introduces the conclusion.

**2. A FEDERATION CONCEPT**

There are real-world applications for the federation concept, including in business and sports.The fundamental concept of federation is the collaboration of multiple independent parties. Federation is not only common in society, but it is also crucial in computing. Federated computing systems have long been a popular field of research in computer science.In the 1990s, federated database systems (FDBSs) were the subject of numerous studies.An FDBS is a group of independent databases that work together for common advantages.As cloud computing has developed, numerous studies on federated cloud computing (FC) have been carried out.Multiple internal and external cloud computing services are deployed and managed under the umbrella of FC. By outsourcing some tasks to more economically advantageous locations, the cloud federation concept enables further cost savings(Li et al., 2021).In 2016, McMahan et al. coined the "Federated Learning" term which refers to a distributed ML technique in which a model is trained collaborativelyby a group of participating clients (such as mobile devices or entire organizations) under the orchestration of a centralized server.Rather than sending or receiving raw data between clients and server, the central server instead receives targeted updates meant for instant aggregation(Kairouz et al., 2021).

## 3. FEDERATED LEARNING WORKFLOWS

In a FL system, the common architecture and training process workflow can be illustrated in figure 1.Two roles often predominate in a FL system : (1)clients that maintain their local data; (2) server that manages the process of training ML model and updating the global model without access to client data.Although there is typically only one server, there may be many clients.In a specific FL systems, a particular client can take on the role of the orchestrating server during the training phase(Yin et al., 2021).

Assume there are *n* clients, each of which is represented by $C_i$, where$i\in$ [1, *n*] and $C_i$ maintains a dataset $D_i$. In a traditional ML approach, all the clients' datasets will be combined in*D*,making $D=D_1\cup...\cup D_n$, and then the ML model will be trained using*D*(Yang et al., 2019).Whereas in a FL approach,a global model is jointly trained by all *n* clients utilizing their local datasets(Yin et al., 2021).

In the FL training process, there are normally three main steps(Yin et al., 2021)(Zhan et al., 2022):

- Initialization: First, the orchestration server determines the global model's architecture,initializes the global model's parameters at random or by pre-training on a public dataset,and initializes the hyper-parameters (e.g., FL rounds number, the total number of clients, and the number of clients chosen for each training round). Then, the server broadcasts the parameters of the initial global model$w_0^g$ to the chosen clients.
- Training local models :In the *t* round, the chosen clients receive the server's most recent global model information (like gradients or weights), and they use their local datasets to update the parameters of their individual local models $w_t^i$.Each participating client will transmit the updated local model parameters $w_{t+1}^i$to the server once the local training is complete.In the *t* round, the client *i* seeks to obtain the ideal local model parametersthrough the reduction of the loss function $F(w_t^i)$,which is expressed by the following equations:

$$w_t^i = \arg min_{w_t^i} F(w_t^i), \qquad (1)$$

$$F(w_t^i) = \frac{1}{|D_i|} \sum_{j \in D_i} f_j(w_t^i), \qquad (2)$$

where $|D_i|$ denotes how many samples are in the dataset $D_i$.Each client's update process can be accomplished by using stochastic gradient descent (SGD) with min-batches sampled from its local dataset:

$$w_t^i = w_t^i - \eta \nabla F(w_t^i), \qquad (3)$$

Where η denotes the learning rate, and $\nabla F(w_t^i)$ is the loss function's gradient.
- Global model aggregation :In each round, the chosen clients' local updated parameters are collected by the server, puts the updated model parameters in lieu of the global model parameters, and then transmits the updated global model parameters$w_{t+1}$ back to the clients for the following training round.In particular, the goal is to obtain the ideal global model parameters$w_t^g$ through the reduction of the global loss function ,which can be described as follows:

$$F(w_t) = \frac{1}{|D|} \sum_{i=1}^{n} |D_i| F(w_t^i), i \in 1,2,\ldots,n \qquad (4)$$

Upon fulfillment of the termination condition (for example, when the number of rounds is increased to its maximum or the global model's accuracy surpasses the threshold), the server terminates the training process, aggregates the updates, and broadcasts the global model to all clients(Yin et al., 2021).

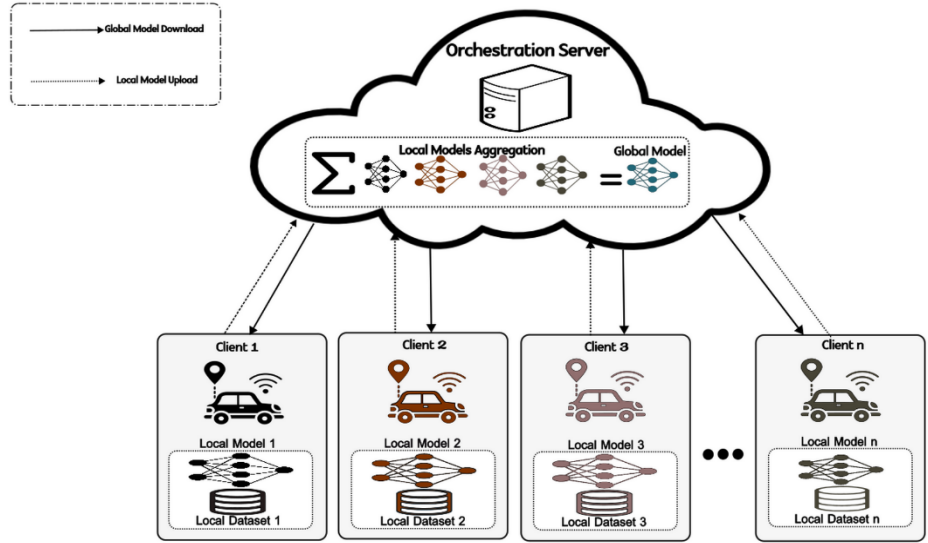

**FIGURE 1.** The Federated Learning Architecture

## 4. FEDERATED LEARNING VS. NON-FEDERATED LEARNING

Traditional centralized-ML approaches build ML models by aggregating distributed raw data generated from various organizations or devices into a single cluster or server(J. Liu et al., 2022)(Tan et al., n.d.). Figure2.(a) depicts a traditional centralized-ML system, which consists of a server and several edge devices. In order to create the centralized training data that the server will use to train the ML model, the edge devices upload their local dataset to the central server. However, centralized-ML approaches encounter critical issues like privacy violations, high communication overheads, and in scalability. To be able to resolve these issues, the idea of FL has been put forth as a practical alternative for traditional centralized ML.as depicted in Figure 2.(b),typically, a FL system consists of a server as well as several edge devices. In a FL, training data is kept locally while multiple edge devices train ML model cooperatively and distributedly under the supervision of a centralized server. In particular, the edge devices implement local training on their own dataset, and they send the results of that training to the server, which update the global ML model after aggregating the received results. The global ML model can attain adequate accuracy after numerous interactions between the edge devices and the server, referring that training is finished(Tan et al., n.d.).

FL differs from traditional centralized-ML approach in a number of important aspects. In FL, Edge devices send model update parameters to the centralized server instead of raw data, which minimizes the communication data size. It thus improves network bandwidth usage. The following aspect is latency which is crucial in time-critical applications including real-time media, industrial control, remote control, and mobility automation. Additionally, real-time decision-making applications like augmented reality and event detection can be executed locally at edge devices to enhance performance. Due to this, FL systems exhibit much reduced latency than centralized-ML systems. The third aspect is privacy: because raw data are not transmitted to a central server, each user's privacy is ensured. Furthermore, FL is simple and uses less power because the models are trained collaboratively on multiple edge devices, implying that edge computing is an appropriate environment for FL (Abreha, 2022).

In general, centralized-ML approaches pose a number of challenges, including training time, computational power, and most significantly, security and privacy of data. Three aspects set FL apart from the centralized-ML approach. First, direct raw data communication is allowed under the centralized-ML approach, but FL forbids it. Second, the centralized-ML approach typically uses one server or cluster in one region owned by a single business whereas FL uses distributed computing resources across several regions or businesses. Thirdly, FL frequently employs encryption or other security measures to ensure security or data privacy, in contrast to the centralized-ML approach, which is less concerned with these security issues(J. Liu et al., 2022).

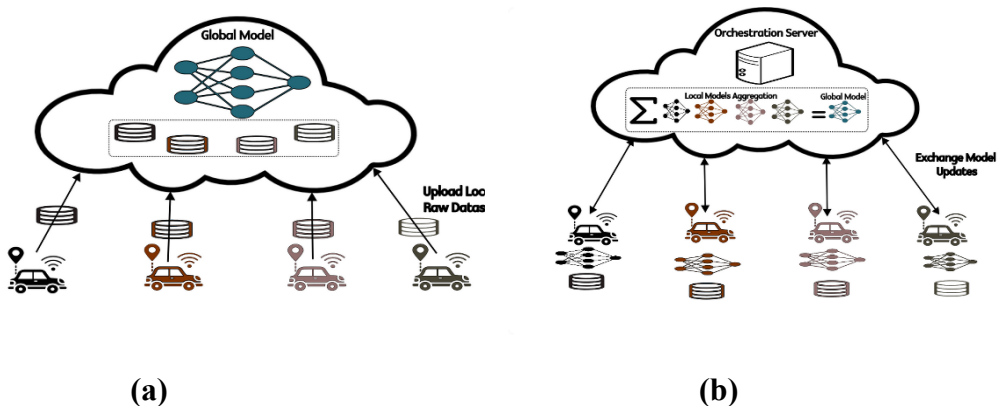

(a)                                   (b)

**FIGURE 2.** General framework for Centralized-ML System& FL System
a) Centralized-ML system (b) FL System.

## 5. SCALE OF FEDERATION
Depending on the federation's scale, the FL systems are divided into two categories: cross-silo and cross-device FL systems. This section will discuss the federation's scale categories.

### 5.1 Cross-silo
Clients in this system are typically small in scale, indexed, and almost always accessible for training rounds, ranging from 2 to 100 devices. Cross-silo FL system is more flexible than cross-device FL system. Clients are typically organizations(Mothukuri et al., 2020). For example, using shopping information gathered from hundreds of data centers worldwide, Amazon wants to recommend products to users. Each data center has ample computational power and access to a vast amount of data(Li et al., 2021).

### 5.2 Cross-device
There are a lot of clients in the cross-device FL system, but each client only has a small data and processing power. Clients are typically mobile devices. One example of a cross-device FL system is Google Keyboard. FL can be used to enhance Google Keyboard's query suggestions(Li et al., 2021).

## 6. FEDERATED LEARNING NETWORK TOPOLOGY

This category focuses on FL's core architecture, or how FL's environment is built from its main component parts. FL can be either centralized or decentralized depending on the network topology as listed below(Mothukuri et al., 2020) :

## 6.1 Centralized-FL

Hub-and-spoke topology, which includes a single server and numerous clients, is considered as the fundamental and most common FL topology which provides a centralized authority in charge of overseeing and managing the training process. Despite the fact that FL is built on the approach of decentralized data nature, still exists a demand for a central server in the FL environment to aggregate trained models from participating clients, create a global model, and sharing it to all FL environment clients (Mothukuri et al., 2020). Cross-device systems typically employ this topology(Kholod et al., 2021).Centralized-FL uses an iterative method called a "federated learning round" that involves multiple client-server exchanges (Shaheen et al., 2022). The following are the primary steps in each round of iterative learning(Kairouz et al., 2021)(Mammen, 2021) **:**

- Client selection: The server chooses participating clients that meet the eligibility criteria.
- Parameter Broadcasting: The chosen participating clients receive a broadcast of the global model parameters from the server.
- Local Model Training: Using their own local datasets, the participating clients will simultaneously retrain their local models.
- Model Aggregation: The server aggregates the participating clients updates. If enough devices have provided results, stragglers can be eliminated for effectiveness.
- Model update: Based on the aggregated update computed from the clients who took part in the current round, the server updates the shared model locally.

The central server requests stopping of the iterative training process when a specified criterion for termination is met. The global trained model is then regarded as a robust model by the FL server(Shaheen et al., 2022).

## 6.2 Decentralized-FL

The peer-to-peer topology is used in the decentralized-FL approach. In this approach, a group of clients can build a ML model and achieve high accuracy without the aid of a third-party centralized server for model aggregation(Mothukuri et al., 2020).This topology is commonly used in cross-silo systems with powerful client computing capabilities(Kholod et al., 2021). Peer-to-peer topology eliminates the central server that could serve as a breeding ground for threats and vulnerabilities. However, the overall FL system may necessitate the involvement of an independent entity in control and arbitration roles, this, because of the increased expenditures and procedural viscosity, isn't always ideal (Bouacida & Mohapatra, 2021).The training process should be orchestrated using a protocol since there is no central server. Generally, there are two protocols to orchestrate the training process(Yin et al., 2021):

- Cyclic transfer: The clients in this protocol are arranged in a circular chain. Client C1 transmits its most recent model information to client C2. Client C2 acquires model information from Client C1 and modifies it using its own data before sending the modified model information to Client C3. The training process is terminated when the termination condition is met.
- Random transfer: In this protocol, a client Ck selects a client Ci at random with an equal probability and sends to Ci its model information. After receiving the model data from Ck, Ci updates it utilizing its own local dataset before selecting client Cj at random and providing the new model information to Cj.

When using networks with high latency or limited bandwidth, it has been demonstrated that decentralized learning is faster than centralized learning. Similar to this, decentralized algorithms in FL have the potential to reduce the central server's high communication costs(Directions & Smith, 2020).

## 7. CLASSIFICATION OF A FEDERATION

FL is often broken down into three categories: vertical, horizontal, and transfer learning depending on the distribution of the data over the sample and feature spaces. These categories are discussed below:

### 7.1 Vertical Federated Learning

Vertical Federated Learning (also referred to as feature-based FL) is used when each participating client has dataset with different features but share the same sample space (Yang et al., 2019).Organizations having data on the same group of people but with varying feature sets can employ vertical FL to build a shared ML model (Mammen, 2021).Vertical FL is suitable for cooperation between the general and specialized hospitals. Despite the fact that they may share the same patient's data, the general hospital is the owner of the patient's generic information and the specialized hospital is the owner of the patient's specific test results. They can therefore utilize Vertical FL to cooperatively train a model that anticipates a specific disease tested by the specialist hospital based on the features of the general hospital (C. Wang & Yao, 2021).

### 7.2 Horizontal Federated Learning

Horizontal Federated Learning (also referred to as sample-based FL) is used when each participating client has the same feature space but different sample instances in their dataset(Yang et al., 2019).Horizontal FLis suitable for credit risk management in banks. Suppose a group of banks all have the same set of features but have a relatively small number of customers. If those banks aggregate their samples, they can create a larger dataset from which to train a more accurate model. However, banks would be wary of disclosing their private information to outside parties for business reasons. Through the use of horizontal federated learning, local models from many banks may be combined to train a global model.

### 7.3 Federated Transfer Learning

Federated Transfer Learning is used when each participating client has a different feature space and sample of data. Traditional ML approaches are typically restricted to resolving issues in one specific domain, i.e., training and testing data must follow the same distribution. Retraining the model on the new dataset is frequently required when the training and testing data's feature distributions are different. However, in real-world scenarios, reacquiring data can be very expensive or even impossible. It's essential to transfer practical knowledge acquired from an existing domain to a new domain without retraining models.Using knowledge from other domains, federated transfer learning can be used to create an efficient model for the target domain(Zhang et al., 2021).

## 8. FEDERATED LEARNING AGGREGATION ALGORITHMS

Several FL algorithms have been proposed for better model updates aggregation. The algorithms for aggregation can be centralized, hierarchical, or decentralized. In what follows, the most popular aggregation algorithms are covered along with a summary of their main merits and demerits, which is shown in Table 1.

**8.1 Centralized Aggregation**

Central servers are used in centralized aggregation algorithms to compute the average models or gradients sent from multiple clients(J. Liu et al., 2022).

Federated stochastic gradient descent (FedSGD) is a standard FL aggregation algorithm built on SGD.FedSGD considers a single SGD step because the chosen participating clients perform one epoch of gradient descent per round. Each round, the participating clients compute the gradients for the local ML models using their own local data and upload the results to the server, the global model is then updated by averaging the gradients from all participating clients (Mcmahan & Ramage, 2017).

The federated averaging algorithm (FedAvg), suggested by McMahan et al. as a more generalized version of FedSGD, is currently the most commonly used aggregation algorithm(Mcmahan & Ramage, 2017).FedAvg allows each client to do an SGD optimization on its local dataset in more than one batch or to run multiple epochs of local training, in this way, less clients-server communication is needed, and instead of exchanging gradients with the server, they do so by exchanging delta weights(Malekijoo et al., 2021)(Kholod et al., 2021).

In this context, a round refers to the uploading procedure of information from the client to the server, updating the global model, and downloading global model information from the server. It differs from an epoch, which is concerned with running the centralized ML model and altering the weights and biases of the local model(Malekijoo et al., 2021).

**8.2 Hierarchical Aggregation**

A hierarchical aggregation is dependent on the use of multiple aggregation servers(J. Liu et al., 2022).

HierFAVG aims to promote partial local models aggregation at the edge-servers level. In this algorithm, there are two different types of communication: client-edge and edge-cloud. In client-edge communication, each edge server aggregates local model updates from its own participating clients, whereas in edge-cloud communication, The aggregated updates from the edge servers are sent to the main cloud server (L. Liu et al., 2020).

Because of the LAN's abundant bandwidth and low financial cost in comparison to the WAN, LanFL algorithm has been proposed to use a hierarchical aggregation mechanism in LAN.LanFL allows for peer-to-peer device organization. A locally shared model is trained by frequently exchanging model updates between devices connected to the same LAN domain. To jointly train a globally shared model, a lot of LAN domains also upload their model updates to a centralized server, but this happens infrequently(Yuan et al., 2020).

**8.3 Decentralized Aggregation**

Traditional centralized FL has a number of issues, including a single-point of failure and the fact that the training process would be abruptly and fully stopped if the central server goes down. In order to handle the transfer of potentially massive volumes of data with all of the clients, the central server also needs to have reliable and high-bandwidth communication links with them. Last but not least, each client needs to have trust in the server(Pappas et al., 2021).Decentralized FL has been introduced in order to address the aforementioned issues. Each device performs the calculation equally under the control of the decentralized aggregation algorithms (J. Liu et al., 2022).

Devices homogeneity is a presumption taken by FL. When FL is used with heterogeneous devices, fast devices waste computing power waiting for slow devices, which is inefficient. The HADFL algorithm has been proposed as a solution to this issue.HADFL enables decentralized peer-to-peer training on heterogeneous devices, allowing these devices to run different local steps in accordance with their computing capabilities(J. Cao et al., 2021).

The IPLS FL algorithm, which is partly based on the interplanetary file system, has been described as a fully decentralized FL algorithm. Each device in IPLS is in charge of some model partitions, and devices communicate with one another by exchanging model updates or partitioned gradients(Pappas et al., 2021).

**TABLE 1** summary of the FL aggregation algorithms

| Algorithm | Type | Merits | Demerits |
| --- | --- | --- | --- |
| FedSGD | Centralized | FedSGD can be performed on resource-constrained devices because it only requires one epoch of SGD implementation in each round. | Since aggregation happens after each local gradient step,this leads to more communications rounds and slowly converges global model training. |
| FedAvg | Centralized | FedAvg, in comparison to FedSGD, reduces the high communication cost and accelerates global model convergence by increasing the number of local training epochs performed on the edge device. | Weights may perform drastically worse when averaged along coordinates, which reduces communication effectiveness. |
| HierFAVG | Hierarchical | In comparison to cloud-based FL, HierFAVG can decrease both the end devices' energy consumption and the model training time. | Due to the close coupling between the two communication types, any failures of an edge server or client device will make the cloud wait incredibly long. |
| LanFL | Hierarchical | LanFL can greatly speed up training and lower traffic on WANs. | LAN introduces additional bias into the device selection process. |
| HADFL | Decentralized | HADFL effectively makes use of heterogeneous computing power while relieving the central server's communication burden. | HADFL's accuracy loss can occasionally be slightly greater than that of other algorithms. |
| IPLS | Decentralized | IPLS is resource-efficient and resistant to sporadic connectivity and dynamic participant arrivals/departures. | There is no incentive mechanism in IPLS. |

## 9. FUNDAMENTALS OF CLOUD, EDGE, AND FOG COMPUTING

The main backbone for any machine or device, including IoT, is computing. As seen in Figure 4, computing may take the form of edge computing, fog computing, or cloud computing (EC) (Sufian et al., 2021) .

Over the last decade, computing paradigms have evolved significantly. Undoubtedly, cloud computing is the most well-known one. This paradigm was developed in response to the need to use "computing as a utility" to make it simple to develop new Internet services. Prior to the IoT revolution, cloud computing was a common research topic; however, the IoT revolution exposed all of the flaws in such a centralized paradigm, igniting interest in decentralized paradigms(De Donno et al., 2019).As a result, fog and edge computing were developed as complements to cloud computing to bridge the cloud-to-things gap by providing a service continuum to a significant number of geographically dispersed IoT devices (Buyya & Srirama, 2019).

In 2006, "Cloud computing" term was first used by Amazon and Google(Kalyani & Collier, 2021).The NIST defines cloud computing as a model for providing practical, global, on-demand network access to a shared pool of reconfigurable computing resources (such as services, storage, servers, networks, and applications) that may be quickly provided and released with little administration work or service provider involvement(De Donno et al., 2019). Even though cloud computing has several advantages, like scalability, effectiveness, cost savings, and reliability, there are some disadvantages when handling large amounts of data. Cloud Computing challenges include real-time analytics, load balancing, high internet bandwidth, latency, energy consumption, data management, security, and privacy. In addition, since it uses a centralized computing model, most operations are performed directly in the cloud. The limitations of cloud computing can be overcome with edge and fog computing (Kalyani & Collier, 2021).

IoT's emergence marked the start of the post-Cloud computing era. The close relationship between cloud and IoT, as represented by directly processing IoT data on the cloud, brings up several issues that cloud computing alone cannot fully address(De Donno et al., 2019). Edge computing is used to meet the industry's requirements for agility in connection, application intelligence, data optimization, real-time business, privacy, and security (K. Cao et al., 2020).Edge Computing, which allows end devices to locally process and store data, has therefore emerged to enhance cloud performance and solve cloud issues(Kalyani & Collier, 2021).The edge computing model does not upload data to a cloud computing platform, instead storing and processing it on edge devices. This characteristic makes edge computing clearly superior in the following areas:(1)Fast data processing: Due to edge computing nodes' closeness to the data source, they can perform computing and data storage tasks locally, reducing the need for intermediate data transmission and minimizing delay time and ensuring real-time processing.(2)Security: All data should be uploaded to the cloud in order to use traditional cloud computing's unified processing feature, which takes a centralized processing approach. This process carries some risks, such as data leakage and data loss . Edge computing ensures data security because there is no need to upload data to the cloud. (3)Low network bandwidth: Edge computing does not require considerable bandwidth on the network due to the fact that processing the data does not necessitate uploading it to a cloud computing (K. Cao et al., 2020) .

Analytically, even when cloud and edge computing are utilized in tandem, some limitations remain because edge nodes are lightweight with limited processing and storage capabilities, resulting in resource contention and increased processing latency. Fog computing has been suggested as a middle resource that can easily combine edge and cloud resources(H. Cao, 2020).At CISCO in 2012, Bonomi coined the term "Fog Computing" for the first time. Fog computing, which puts processing and intelligence near to the data sources, expands on the cloud computing architecture(Idrees & Idrees, 2022). Fog computing offers a new layer of computing that sits between conventional cloud computing and distributed IoT devices (Tmamna et al., 2020). As a result, the fog computing layer's data reduction can result in a faster response time to smart end devices while also lowering the size of the data uploaded to the cloud platform, resulting in network bandwidth savings(Idrees & Idrees, 2022).Fog computing and edge computing share many of the same principles such as mobility support, low bandwidth costs, low latency, high scalability, and virtualization service. It does, however, have constrained resources, fewer computational and storage options, and is closer to end devices than Fog computing(Kalyani & Collier, 2021).

In particular, Fog and Edge Computing offers five key benefits, which include: (1)Security: Fog and Edge Computing support extra security for IoT devices to guarantee transaction security and reliability; (2)Cognitive: Fog and Edge Computing make it possible for their clients to be aware of their goals in order to support autonomous decision-making about when and where to deploy storage, computing, and control functions; (3) Agility: Fog and Edge Computing improve the agility of the widespread deployment of an IoT system;

(4)Latency: Fog and Edge Computing offer fast responses for applications that demand extremely low latency; (5) Efficiency: Fog and Edge Computing enhance cloud computing efficiency by enhancing performance and cutting unnecessary costs (Buyya & Srirama, 2019).Figure 3 shows the cloud server, fog gateway, and edge devices in a hierarchical relationship.

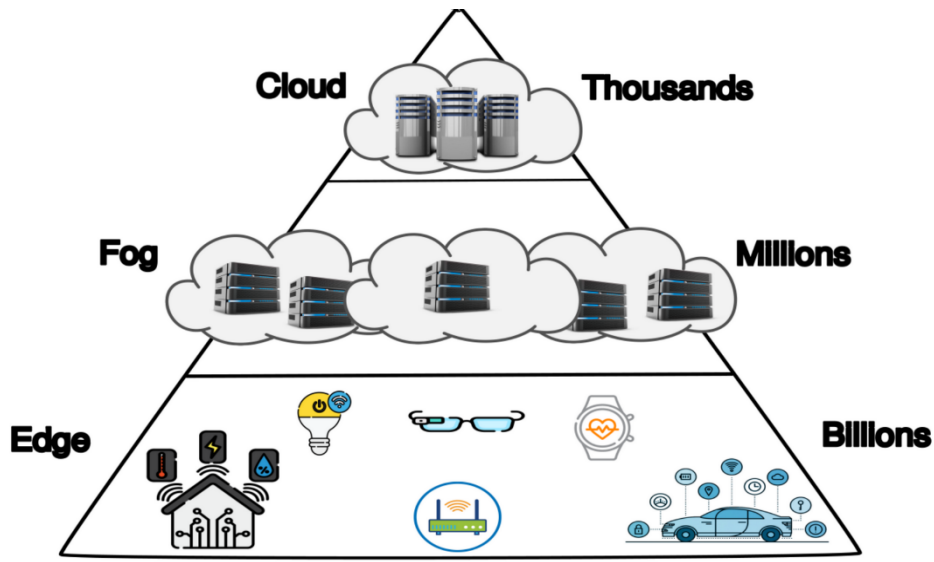

**FIGURE 3.** Cloud, Fog, and Edge Computing Hierarchy

### 10. ENABLING DEEP LEARNING ATTHE EDGE/FOG

Machine learning (ML), which has been around since the 1950s, has revolutionized a number of fields in recent years. Deep learning (DL) emerged primarily after 2006 from a branch of machine learning (ML) known as neural networks (NN) (Alom et al., 2019). DL is also known as representation learning (RL) (Alzubaidi et al., 2021). Since its inception, DL has demonstrated exceptional success in almost every application domain (Alom et al., 2019).DL draws inspiration from how the human brain processes information. DL doesn't need any rules that were created by humans to work; instead, It makes extensive use of data to match the input to specific labels (Alzubaidi et al., 2021).Using traditional ML techniques, the classification task must be completed in a series of actions that happen one after another, such as preprocessing, feature extraction, feature selection, learning, and classification. Moreover, the performance of ML techniques is significantly impacted by feature selection. Biased feature selection may result in incorrect class distinction. In contrast, unlike conventional ML techniques, DL can automatically learn feature sets for a variety of tasks (Alzubaidi et al., 2021).

DL is defined as the use of multiple-neuron, multiple-layer neural networks to carry out learning processes like clustering, classification, and others(Hatcher & Yu, 2018).A basic neural network is made up of three layers: input, hidden, and output. In a simple neural network, no computation occurs in any of the input nodes; instead, the information is passed on to the hidden nodes. As a result, this network only has two layers. Traditional computing systems based on central processing units are insufficient to support multilayer architectures because there are thousands of interconnections between these layers in deep architectures. By using GPU computing and having access to a lot of training data, it was possible to add additional layers between the input and output layers. A simple neural network is changed into a deep neural network by including two or more hidden layers (Khamparia & Singh, 2019).Figure 4 depicts a simple neural network and a DNN with two hidden layers.

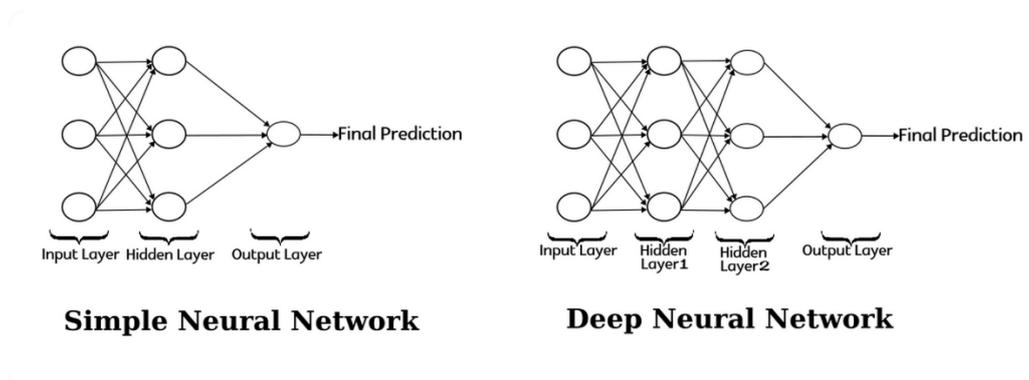

**FIGURE 4.**Neural Network VS Deep Neural Network(Khamparia & Singh, 2019)

Applications of DL can be found in many aspects of our lives, including image recognition, image description or caption generation, speech recognition, demographic prediction, earthquake prediction, and election result prediction (Khamparia & Singh, 2019).

Depending on the training approach, DL can be classified into one of two categories: Centralized and Federated Training. In a traditional centralized training approach, the server requests that clients send their training samples to the cloud. After initializing DNN in the cloud, the central server uses training samples to train the DNN until the ideal parameters are found. Finally, the cloud server gives the users the ideal parameters. In FL, as opposed to centralized training, in order to train a single DL model, each user and server cooperate with each other. The cloud server receives local parameters (i.e., gradients) from each user, aggregates them, and then sends the results to each user, accelerating the model's convergence. The best network parameters will ultimately be used by the server and each user(Xu et al., 2019).

Recently, IoT has become more prevalent in several fields including health, transportation, agriculture, education, and smart cities. The majority of these fields' applications depend on an intelligent learning strategy for data mining, pattern recognition, prediction, or data analytics. In recent years, DL has been commonly used in a variety of IoT applications (Mohammadi et al., 2018).So far, DL models are primarily trained on computing platforms that are powerful, such as a cloud datacenter, using massively collected centralized datasets. However, since many applications generate and distribute data at network edge devices like sensors and smart-phones, transferring those data to a centralized server for model training would poses several challenges (L. Liu et al., 2020) such as:
- Latency: Sending data to the cloud for training cannot satisfy the stringent low-latency criteria required for real-time applications.
- Scalability: Scalability issues arise when data is moved from sources to the cloud because as more devices are connected, network access to the cloud may become more congested.
- Privacy: Transferring data to the cloud exposes users who own the data to potential privacy issues(Chen & Ran, 2019).To address the preceding issues, distributed training has begun to draw a lot of attention. Google suggested using FL to train a DNN without having the data stored in a central server(L. Liu et al., 2020).

DL has gradually shifted towards network edges and fog computing as new computing paradigms to improve decision-making quality while addressing the aforementioned latency, scalability, and privacy issues. Because IoT data is generated close to the end user, the newly emerging concept of DL at network edges and fog computing is thought to be a better solution. It has built-in advantages in a wide range of industries, including healthcare, self-driving cars, smart cities, and numerous other applications (Tmamna et al., 2020).

Pushing DL processing from the cloud to the edge has a number of benefits, such as

- The cost and latency of data transmission for processing are greatly reduced due to the deployment of DL services near to the users that are submitting requests, and the cloud only takes part when more computation is needed.
- User privacy is better protected because DL services require local storage of the raw data on the edge rather than cloud storage.
- A more trustworthy DL computation is offered by the hierarchical computing architecture.
- Edge computing, which makes use of more diverse data and application scenarios, can allow the widespread use of DL and achieve the potential of "offering AI for every individual and every company at everywhere".
- Diverse and beneficial DL services can increase edge computing's commercial value and hasten its adoption and expansion(X. Wang et al., 2020).

## 11. FEDERATED LEARNING IMPLEMENTATION IN EDGE/FOG COMPUTING

Numerous tools and frameworks are now available as a result of the rising demand for FL technology. The most popular FL frameworks will be introduced in this section:

- TensorFlow Federated (TFF):The open-source FL framework TFF is built on top of TensorFlow. Google developed TFF in 2019 to allow developers and researchers to simulate federated learning scenarios using TFF's built-in FL algorithms, as well as to allow developers to create their own new FL algorithms. However, the TFF framework does not support the implementation of true federated learning on real devices (TensorFlow, 2022).
- Federated AI Technology Enabler (FATE): FATE is an open-source FL computing framework launched in January 2019. In order to implement secure computation protocols, it makes use of homomorphic encryption and multi-party computation. Developers can deploy FATE on a single node for simple testing or on multiple nodes for scalability, reliability, and manageability (Webank, 2019).
- Paddle Federated Learning (PFL):Baidu scientists developed PFL in 2020 based on PArallel Distributed Deep LEarning (Paddle).PFL supports both vertical and horizontal FL. Developers can use PFL to deploy FL systems on massively distributed clusters (PaddlePaddle, 2022).
- PySyft: PySyft, a Python library developed by OpenMined, uses the concepts of Encrypted Computation, Differential Privacy, and Federated Learning to decouple the training of global model from the private data. It can be used within TensorFlow and PyTorch, two popular deep learning frameworks (Ryffel et al., 2018).
- Flower (flwr): flwr was developed by Daniel et al. as a unified solution for federated learning, analytics, and evaluation.flwr was built to support federated learning at large-scale real-world systems. flwr is compatible with a variety of operating systems and hardware platforms, allowing it to function effectively in heterogeneous edge device environments (Beutel et al., 2022).
- NVIDIA Federated Learning Application Runtime Environment (NVIDIA FLARE): NVIDIA FLARE is an open-source software development kit that enables federated learning among various parties. It includes privacy-protecting algorithms that ensure that the global model's changes are all hidden and guard against the server's ability to reverse-engineer the submitted weights(*NVIDIA FLARE*, n.d.).

## 12. FL DATASETS

There were several datasets available for the implementation of FL. While some were open to the public, others weren't(Shaheen et al., 2022).For instance,LEAF was one of the first federated learning dataset proposals. It includes six datasets from various domains, such as image classification, next-character prediction and sentiment analysis(Li et al., 2021). Shakespeare dataset which is based on each speaking role in William Shakespeare's Complete Works,Federated Extended MNIST (FEMNIST) addresses natural heterogeneity caused by writing style,StackOverflow dataset which is comprised of StackOverflow data that is maintained by the Google team(J. Liu et al., 2022).

## 13. FEDERATED LEARNING IN EDGE/FOG COMPUTING: CASE STUDIES

One of the newest and most exciting technologies is FL. When protecting privacy and making the best use of available resources are priorities, FL on E/F computing is widely applicable. This section will introduce a few use cases for implementing FL in E/F computing as well as some recent research that was applied to them.

### 13.1 Healthcare Systems

Although the edge node stores and processes each medical institution's data separately, the model trained on a tiny dataset from a single medical institution performs poorly when it is applied to unseen data. Consequently, a substantial number of actual electronic health records are necessary to train a robust medical model. Due to the privacy and sensitivity of medical data, it is challenging to meet the demand for real datasets. Institutions in the medical field can work together on training models using FL on edge computing to resolve this problem and adhere to the demands of the Health Insurance Portability and Accountability Act without disclosing patient data(Xia et al., 2021)**.**

Hakak et al. (Hakak & Ray, 2020)proposed personalized Edge-based Federated Learning approach to assist healthcare practitioners by providing insights for disease diagnosis and prognosis via distributed wearable devices data analysis. The importance of incorporating FL into system architecture stems from the data privacy of sensitive medical data generated via wearable devices, as well as ensuring fast response time.

Qayyam et al. (Qayyum et al., n.d.)proposed a clustered FL model that can diagnose COVID-19 based on the classification of images of the chest from different sources, including X-rays and ultrasound, without disclosing any information regarding the training data samples. With the help of this model, remote healthcare centers without cutting-edge diagnostic equipment can take advantage of the collaborative learning paradigm. A collaborative ML model can be built in a cloud-edge infrastructure using the visual data (such as X-rays and ultrasounds) that is gathered at various healthcare facilities. The edges of the cluster represent healthcare organizations (remote medical imaging facilities), whereas the cloud server represents large hospitals or other governmental bodies (like the ministry of health) which facilitates the weights aggregation.

## 13.2 Vehicular Networks

Vehicular edge computing (VEC), a field of vehicular technology that is quickly developing, uses moving vehicles and roadside servers at the network's edge to offer storage, computation, communication, and data resources to other nearby vehicular users. So as to fulfill the growing technical demands of AI applications in vehicular networks, it is urgent that FL in VEC be incorporated as a key technical approach(Ye et al., 2020).

Ye et al.(Ye et al., 2020) proposed a method for FL in VEC for image classification using selective model aggregation. They assessed image quality in terms of motion blur using a geometric model. A resource consumption parameter is then used to quantify the computational ability. After assessing quality of local images and computation ability, the "fine" participating clients' DNN models are chosen and transferred to the main server for aggregation.

Lu et al.(Lu et al., 2020) proposed VEC asynchronous FL scheme. Local differential privacy was incorporated into the gradient descent local training process to keep each participant client's local models secure. They proposed replacing the conventional update system between clients and a centralized server with a random peer-to-peer update mechanism because the centralized curator raised security and privacy concerns.

## 13.3 Unmanned Aerial Vehicles (UAVs)

UAVs are gaining popularity because they can perform a variety of difficult tasks in the three-dimensional space (Mowla et al., 2020).

The ability to control a large number of UAVs in real time enables mission-critical applications such as search-and-rescue missions to deliver first-aid kits, covering large disaster sites, and firefighting scenarios. UAVs are very difficult to control in windy conditions, which increases the risk of collisions and slows down travel times. Mean-field game (MFG) has been used for massive UAV control to restrict inter-communications between UAVs by resolving two stochastic differential equations that are coupled: the Hamilton-Jacobi-Bellman (HJB) equation for optimal control decision and the Fokker-Plank Kolmogorov (FPK) equation for the population distribution estimation. To estimate the HJB and FPK equation solutions, each UAV uses two DNNs. Federated learning is used to ensure convergence and expedite DNN training by periodically exchanging the DNNs' parameter sets for solving the HJB and FPK equations. MFG control based on federated learning can cut travel distance and collision risk by half (Shiri et al., 2020).

A jamming attack, in essence, obstructs devices communication by damaging communications reception at the receiver with the least amount of transmission power. To aid a defense strategy based on reinforcement learning, Mowla et al. proposed a defense strategy-based FL that allows for the detection of on-device jamming attack. Input from the FL model is used in the suggested defense strategy to update a Q-table via the Bellman equation. Furthermore, an adaptive epsilon-greedy policy is used over a combined FL and reinforcement learning model to select the best UAV defense paths. As a result, the UAVs successfully avoid jamming areas that the federated learning approach has identified in advance (Mowla et al., 2020).

## 13.4 Smart City

Smart cities offer reliable solutions to crucial issues with traffic, healthcare, education, security, and other areas. For a variety of uses, smart cities make extensive utilization of intelligent IoT devices (Albaseer et al., 2020).

for developing a framework for managing video data in smart cities, Chiu et al. (Chiu et al., 2020) proposed an edge learning system that incorporates FL and semi-supervised learning. Live street videos can be gathered from connected vehicles' cameras, which are then sent to edge devices. Following that, a semi-supervised learning algorithm is run on each edge device in order to perform local video analytics on a variety of video segments. A "Federated Swapping" operation is suggested as a solution to the non-IID data issue in order to decrease data diversity and improve image classification accuracy.

Albaseer et al. (Albaseer et al., 2020) proposed the FedSem technique of semi-supervised FL that uses unlabeled data in smart cities. Their system consists of a number of autonomous vehicles traveling along various highways. These vehicles are controlled by a central edge server, which monitors the training process. Using the vehicle's own data (images of traffic signs), each vehicle locally trains the local DNN model, then the central edge server receives just gradients from edge devices. In each learning round for model orchestration, multiple participants will then be chosen under the supervision of a central server.

### 13.5 Human Activity Recognition (HAR)

HAR aims to identify a person's activities based on observations about humans and their surroundings. Application areas for HAR include healthcare, behavior analysis, electroencephalography, surveillance, and gesture recognition (Xiao et al., 2021).

Zhao et al.(Zhao et al., 2020) proposed a system that uses FL to help with training and edge devices to carry out local health and activity monitoring. The proposed system gathers and manages sensor and IoT data using the Databox platform, then uses this data to deduce activities locally on an edge device, allowing for the realization of health and activity monitoring. It makes use of a cloud server to manage various devices and conduct global LSTM model training without revealing their raw data.

Xiao et al. (Xiao et al., 2021) proposed HARFLS, a FL system for HAR using wearable-based sensors. It extracts features with a perceptive extraction network (PEN) and shares model weights with the Federated Averaging method. Furthermore, to lower the risk of data leakage during the distribution and uploading of weights, homomorphic encryption is used.

### 14. OPEN ISSUES AND FUTURE RESEARCH DIRECTIONS

The future research directions that we think would be exciting to work on and explore are discussed in this section.

### 14.1 Intelligent Brokers

In a traditional hierarchical FL system, the function of intermediate aggregation servers lacks intelligent decision-making because they only serve as brokers between the central server and the edge devices. Therefore, we highlight the importance of implementing an intelligent broker in a hierarchical FL architecture. The intelligent brokers should make critical decisions before distributing the central server's global model update to edge devices and before forwarding edge devices' local model updates to the central server. For instance, rather than selecting edge devices at random to take part in the training process, they can select the most suitable edge devices based on their real-time computing resource information, dataset size, and historical data such as the accuracy of their local model updates.

Intelligent brokers can also spot edge devices that are submitting local updates that aren't relevant and prevent uploading their updates to the central server.

### 14.2 Stragglers Identification Module

Despite the fact that FL has been applied to a heterogeneous system, it is unable to efficiently utilize the system's heterogeneous resources. The performance of the FL training process on heterogeneous systems is significantly impacted by stragglers since they take an inordinate amount of time to report local updates. As a result, we believe it will be crucial to establish a stragglers identification and mitigation module as a key enabler for heterogeneous FL systems. This module should be able to identify the causes of stragglers' existence, such as data skew, failures, and resource contention, and then it should try to mitigate the effect of the stragglers, for example, it might mark straggler nodes as deprecated when the central server collects local model updates.

### 14.3 Containerized FL

The variety of hardware devices employed in edge and IoT ecosystems poses a major challenge. These devices differ in terms of their sensors, computing power, and cost of network connectivity. Consequently, it is challenging to deploy local ML models dynamically to edge and IoT devices. Since we can remotely create, upgrade, and destroy applications in an elastic manner with minimum overhead thanks to containerization technology, we fervently support the need to dynamically deploy and manage ML models on diverse hardware devices through containerization technology.

### 14.4 Intelligent single-point of failure prediction

In the vast majority of existing FL systems, the centralized architecture is extensively adopted. This architecture may result in a number of severe issues, such as a single point of failure. As a result, we highlight the importance of intelligently predicting central server failures using ML methodologies prior to failure occurrence.

### 14.5 Developing client-specific ML model

The edge computing network is more diverse than traditional central networks, with clients differing based on compute, storage, and network resources, as well as data collection resources. Therefore, client-specific ML model should be developed to deal with such heterogeneities.

## 15. CONCLUSIONS

FL permits distributed clients' devices to jointly train a global ML model while preserving all training dataset on their devices, relieving ML from the requirement to maintain data on the cloud. This extends beyond simply utilization of local models on mobile devices by additionally transferring model training to the device. This chapter introduces a review of the application of FL in E/F computing systems. The Federation Concept, FL Workflows, Federated Learning vs. Non-Federated Learning, Scale of Federation, FL Network Topology, and Classification of a Federation are presented and explained. We look into Federated Learning Aggregation Algorithms. The fundamentals of cloud, fog, and edge computing are introduced. Then, enabling Deep Learning at the E/F is presented. The Investigations are being conducted into a number of case studies involving the application of federated learning in E/F computing. The unresolved problems and potential research directions are presented.

**KEYWORDS**

Federated Learning,
IoT,
Edge/Fog computing,
Deep Learning,
Federation.